\documentclass[twocolumn,english,aps,prb,showpacs,byrevtex,amsmath,amssymb,superscriptaddress]{revtex4-1}

\usepackage{graphicx}
\usepackage{bm}
\usepackage{gensymb}
\usepackage{upgreek}
\usepackage{hyperref}
\usepackage{epstopdf}
\usepackage{dcolumn}
\usepackage{booktabs}
\usepackage{multirow}
\usepackage{hyperref}

\usepackage{amssymb,mathtools}
\usepackage{color}
\usepackage{amsmath} 
\usepackage{physics}

\usepackage{tikz,xcolor,hyperref}
\definecolor{lime}{HTML}{A6CE39}
\DeclareRobustCommand{\orcidicon}{%
	\begin{tikzpicture}
	\draw[lime, fill=lime] (0,0)
	circle [radius=0.16]
	node[white] {{\fontfamily{qag}\selectfont \tiny ID}};
	\draw[white, fill=white] (-0.0625,0.095)
	circle [radius=0.007];
	\end{tikzpicture}
	\hspace{-2mm}
}

\foreach \x in {A, ..., Z}{%
	\expandafter\xdef\csname orcid\x\endcsname{\noexpand\href{https://orcid.org/\csname orcidauthor\x\endcsname}{\noexpand\orcidicon}}
}

\begin{document}

\title{Topological phase diagram of  Pb$_{1-x}$Sn$_x$Se$_{1-y}$Te$_y$}

\author{Giuseppe Cuono\orcidA}
\email{gcuono@magtop.ifpan.edu.pl}
\affiliation{International Research Centre MagTop, Institute of Physics, Polish Academy of Sciences, Aleja Lotnik\'ow 32/46, PL-02668 Warsaw, Poland}
\author{Ghulam Hussain\orcidB}
\affiliation{International Research Centre MagTop, Institute of Physics, Polish Academy of Sciences, Aleja Lotnik\'ow 32/46, PL-02668 Warsaw, Poland}
\author{Amar Fakhredine\orcidE}
\affiliation{Institute of Physics, Polish Academy of Sciences, Aleja Lotnik\'ow 32/46, PL-02668 Warsaw, Poland}
\author{Carmine Autieri\orcidD}
\affiliation{International Research Centre MagTop, Institute of Physics, Polish Academy of Sciences, Aleja Lotnik\'ow 32/46, PL-02668 Warsaw, Poland}

\begin{abstract}
We reproduce the mirror Chern number phase diagram for the quaternary compound Pb$_{1-x}$Sn$_x$Se$_{1-y}$Te$_y$ combining accurate density functional theory and tight-biding model. The tight-binding models are extracted from {\it ab-initio} results, adding constraints to reproduce the experimental results. By using the virtual crystalline approximation, we calculated the mirror Chern number as a function of the concentrations $x$ and $y$.
We report different hamiltonians depending on whether we want to focus on the experimental band gap or on the experimental topological transition.
We calculate the transition line between the trivial insulating phase and topological insulating phase that results in good agreement with the experimental results. Finally, we add in the phase diagram the Weyl phase predicted in the literature providing a complete topological phase diagram for the Pb$_{1-x}$Sn$_x$Se$_{1-y}$Te$_y$ quaternary compound.
\end{abstract}

\date{\today}

\maketitle

\section{Introduction}

In the last decade, there was a new interest in the investigation of narrow gap semiconductors and semimetals due to the chance to engineering 
magnetism\cite{Dietl_2019,RevModPhys.87.1311,Dietl2021,Kepa03,Humpfner07}, topological insulator phases\cite{HgTe_weyl,PhysRevB.104.L220404}, Dirac and Weyl phases\cite{Qian20Weyl,PhysRevResearch.4.023114,Campbell2021,PhysRevB.105.235304}, quantum spin Hall phases\cite{Konig:2007_S}, nodal-line phases\cite{PhysRevX.6.031003,PhysRevB.101.205149}, axion insulating phase\cite{Pournaghavi2021Realization,Islam2022MnTe} and flat bands\cite{lau2021designing}. Indeed, heterostructures and doping allow to manipulate the electronic, optical and magnetic properties producing new phases\cite{vanthiel2020coupling,PhysRevLett.127.127202,Autieri2016,Autieri2014NJP,Amitesh2014APL,PhysRevB.85.075126,PhysRevResearch.4.023256}.
Among these phases, the topological crystalline insulators (TCI) are a state of matter in which the topological nature of the electronic structure arises from the crystal lattice \cite{hsieh2012topological}.
The surface bands of TCI are low-energy states occurring
along certain high-symmetry directions \cite{Barone13a,Barone13,Volobuev17}.
The topological surface states in TCI are protected by crystalline space group symmetries, instead of the time-reversal symmetry as in the case of the conventional Z$_2$ topological insulators (TI) \cite{Fu07}.
The topological crystalline phase has been found also in semimetallic systems \cite{Sun20,Wadge22}, while topological crystalline metallic phases have been deeply investigated mainly in orthorhombic crystal structure \cite{Chen15,PhysRevB.97.081105,Daido19,Autieri17a,Autieri17,Autieri18,Cuono19,Cuono19b,Nigro22} as well as in other transition metal pnictides with quasi-one-dimensional structure \cite{Xu20,Cuono19c,Cuono21,Cuono21b} where the coexistence of topology and superconductivity has been predicted.
\\

IV-VI semiconductors \cite{Dietl21} exhibit many properties such as thermoelectricity \cite{wang2011heavily, wood1988materials}, ferroelectricity \cite{lebedev1994ferroelectric, liu2020synthesis}, superconductivity \cite{matsushita2006type, mazur2019experimental} and zero-energy modes \cite{Sessi16,Brzezicki19}.
These compounds are used for applications in spintronics \cite{Jin09,Grabecki07} and optoelectronics \cite{Akimov93,Liu10}.
Much of the interest in these materials is due to the recent discovery of topological crystalline insulating phase in SnTe \cite{fu2011topological, hsieh2012topological,Lau19} and some of their substitutional alloys \cite{xu2012observation,dziawa2012topological,disante15} such as Pb$_{1-x}$Sn$_x$Te \cite{xu2012observation}, Pb$_{1-x}$Sn$_x$Se \cite{dziawa2012topological} and Pb$_{1-x-y}$Sn$_{x}$Mn$_{y}$Te \cite{lusakowski22}.
While SnTe is a cubic TCI, it was shown that SnSe is an orthorhombic topological system \cite{PhysRevMaterials.6.054201}.
Furthermore, it has been shown that SnTe is helical higher-order topological insulator \cite{Schindler18,Kooi20,vanMiert18}. It has been observed that the characteristic properties change by changing the size or dimensions (1D, 2D or 3D phase) of materials, namely the properties will change if we move from bulk \cite{Plekhanov14,Wang20,Barone13} to thin films \cite{Slawinska20,Volobuev17,Liu18,Galeeva21,Kazakov21,Rechcinski21} and to nanowires\cite{Nguyen21,Hussain22_nanowires}. For instance, SnTe is trivial insulator at low thickness but becomes topological above some critical thickness\cite{liu2014spin, liu2015crystal}. 
In previous works, it has been shown that another way to tune the topological properties is to investigate how topology changes as a function of the composition in Pb$_{1-x}$Sn$_x$Te and Pb$_{1-x}$Sn$_x$Se \cite{dziawa2012topological,Tanaka12,xu2012observation,Wojek14,Islam19}, in fact, PbSe and PbTe are trivial, while SnSe and SnTe are TCI.
In the intermediate region between the trivial and the TCI phase for the rock-salt chalcogenides alloys, the presence of a Weyl phase (WP) has been proposed \cite{Lusakowski18,Wang19b}.\\ 

In this paper, we will investigate how the doping can tune the different topological properties as the trivial insulating phase, the topological crystalline phase and the Weyl phase in the quaternary compound of the IV-VI semiconductors Pb$_{1-x}$Sn$_x$Se$_{1-y}$Te$_y$.
For the first time, we analyze the properties of the quaternary compound Pb$_{1-x}$Sn$_x$Se$_{1-y}$Te$_y$ as a function of the concentrations x and y by using the Virtual Crystal Approximation (VCA).
We use a density functional theory (DFT) approach and we build tight-binding minimal models based on the Wannier transformation of the {\it ab-initio} results.
We report different tight-binding models depending on whether we want to reproduce the gap or the Chern number.
We look at the concentration which leads to the transition from a topologically trivial phase to a non-trivial phase. 
With our models, we reproduce reasonably well the experimental results and give an important indication about the transition line from the trivial to the topological phase. 
The WP phase predicted in the literature is located in a region of 15\% of doping concentration starting from the point of the closing of the gap in the VCA approximation\cite{Lusakowski18,Wang19b}.
The paper is organized as follow: in the next section we describe the computational details, in the Sec. III the results are reported, with three different Chern number maps obtained with three sets of parameters, while the last section is devoted to the conclusions.

\section{Computational details}
We performed our calculations within the framework of the first-principles density functional theory (DFT) based on plane-wave basis set and projector augmented wave method using VASP \cite{VASP} package. The calculation is fully relativistic by considering spin-orbit coupling (SOC). A plane-wave energy cut-off of 250~eV has been used.  We have performed the calculations using 6$\times$6$\times$6 k-points centered in $\Gamma$ with 216 k-points in the independent Brillouin zone. 
\\
For this class of compounds, the generalised gradient approximation (GGA)  \cite{perdew1996generalized} underestimates the band gap. By using GGA we obtain that the compounds are topological for all the concentrations. Several methods have been used in the literature in order to increase the gap, including the GGA + U with Coulomb repulsion on the 6$s$ orbital of Pb \cite{Wang19b}. 
Instead, we have used the meta-GGA approach the modified Becke-Johnson (MBJ) \cite{Goyal17,MBJLDA1} exchange potential together
with GGA for the correlation potential scheme, as already done in previous works for SnTe material class\cite{MBJLDA,Islam19} and other narrow gap semiconductors\cite{PhysRevB.103.115209,Hussain22}.
We used a value of the parameter c=1.10 that gives us good results for the band gap and for the mirror Chern number.
After obtaining the Bloch wave functions $\psi_{n,\textbf{k}}$, the p-like anion and cation Wannier functions are build-up using the WANNIER90 code\cite{Mostofi08}. To determine the real space Hamiltonian in the Wannier function basis, we have used the Slater-Koster interpolation scheme, and  we have constructed the symmetrized relativistic Wannier tight-binding model using an independent python package wannhrsymm\cite{WANNSYM}. 

We have used the room temperature lattice constants\cite{McCann87} reduced by 0.5\% in order to consider the temperature effect. The lattice constant used for the zero kelvin DFT calculations are: $a_{SnTe}$=6.2964~\AA, $a_{PbTe}$=6.4277~\AA ~  and $a_{PbSe}$=6.0954~\AA. 

\section{Mirror Chern number calculation}

We used the Virtual Crystal Approximation (VCA) to estimate the transition between topologically trivial and non-trivial phases.
We analysed the trend of the Mirror Chern Number (MCN)  for the quaternary compound Pb$_{1-x}$Sn$_x$Se$_{1-y}$Te$_y$ as a function of the concentrations x and y. Then, we studied the transition from the trivial to the topological phase as a function of x and y.
The VCA Hamiltonian that describes the alloy is:

\begin{equation}
\label{eqn:alloy}
H(x,y)=xH_{SnTe} + (y-x) H_{PbTe}+ (1-y)H_{PbSe} \, ,
\end{equation}
where $H_{PbTe}$, $H_{PbSe}$ and $H_{SnTe}$ are the symmetrized relativistic Wannier tight-binding Hamiltonians for the PbTe, PbSe and SnTe compounds, respectively.
We have excluded the SnSe from equation (\ref{eqn:alloy}) because it has a different structure, rhombohedral instead of cubic, and it is far from the transition. 
We note that the Hamiltonian exhibits the mirror and the time-reversal symmetries.
The mirror operator with respect to the ($\overline{1}$10) plane is:

\begin{equation}
\label{eqn:mirror}
M_{xy}=\frac{1}{\sqrt{2}}(\sigma_{x}-\sigma_{y})
\otimes(\mathbb{I}_{3}-L_{z}^{2}-\comm{L_{x}}{L_{y}})\otimes P_{xy}\,.
\end{equation}

It is a Kronecker product of three parts related to the spin, the orbital and the  atomic degrees of freedom.
$\sigma_{x}$ and $\sigma_{y}$ are the Pauli matrices, $\mathbb{I}_{n}$ is the n$\times$n identity matrix, $L_{i}$ with $i=x,y,z$ are the orbital angular momentum operators for $l=1$, and $P_{xy}$ is the matrix that exchanges the atomic positions respect to the mirror plane, whose form depends on our choice of the vector basis.
The time-reversal operator is $T=\sigma_{y}\otimes \mathbb{I}_{3}\otimes \mathbb{I}_{8},$  also product of three parts, the first related to the spin, the second to the orbital and the third to the atoms.
These operators verify the following relations:
\begin{equation}
\label{eqn:actionmirror}
M_{xy}H(k_{x},k_{y},k_{z})M^{\dagger}_{xy}=H(k_{y},k_{x},k_{z})\, ,
\end{equation}
\begin{equation}
\label{eqn:actiontime}
TH(k_{x},k_{y},k_{z}){T}^{\dagger}=H(-k_{x},-k_{y},-k_{z})\, ,
\end{equation}
\noindent and the mirror operator anticommutes with the time-reversal operator. In the eigenbasis of the mirror operator, the time-reversal takes off-diagonal block and the Hamiltonian commutes with the mirror operator. Therefore, it has a block-diagonal form. This leads the MCN vanishing on the high symmetry planes, where each block has an opposite MCN. 
To calculate the MCN that from now on we define as $C$, we use the Kubo formula:
\begin{equation}
\label{chernnumber}
\footnotesize C=\frac{1}{\pi}\int_{0}^{2\pi}\int_{0}^{2\pi}\sum_{n\leq n_{F}, n'>n_{F} } Im\left[\frac{\small \langle n \small \lvert \partial_{k_{x}}H \rvert n' \small  \rangle \small \langle n' \small \lvert \partial_{k_{z}}H  \rvert  n \small \rangle}{(\epsilon_{n}-\epsilon_{n'})^2}\right]dk_{x}dk_{z}\, ,
\end{equation}
where $ \rvert n \small \rangle $ and $\epsilon_{n}$ are the eigenstates and the eigenvalues of the Hamiltonian in the projected subspace and $n_{F}$ is the filling.
When the $C$=0, we have a trivial insulator, while when it is equal to 2 we have a TCI.

\section{Realistic and transferable tight-binding for P\lowercase{b}S\lowercase{e}, P\lowercase{b}T\lowercase{e} and S\lowercase{n}T\lowercase{e}} 

Here, we report the minimal tight-binding models for the selected compounds that reproduce reasonably well the properties of the quaternary compound Pb$_{1-x}$Sn$_x$Se$_{1-y}$Te$_y$ as a function of the concentrations x and y.
It is not possible to fit both the gap and the closure of the trivial band gap with a simplified tight-binding hamiltonian. Therefore each model here presented focuses on one property that we want to reproduce.
We report a model which reproduces the experimental band gaps at L and at T=0 for PbTe and PbSe and the theoretical band gap for SnTe in Table \ref{tabhoppings1}, a hamiltonian which fits the closure of the trivial band gap in Table \ref{tabhoppings2}, and a model which reproduces the experimental band gaps at T=0 for PbTe, PbSe and SnTe in Table \ref{tabhoppings3}.\\

In the Tables, we show the on-site energies  $\epsilon$, the first and the second neighbour hopping parameters and the spin-orbit coupling $\lambda$ constants of our tight-binding models for SnTe, PbTe and PbSe\cite{Islam19}.
The first neighbour hopping is the cation-anion hopping; the second ones are anion-anion (aa) and cation-cation (cc) hopping terms. 
We denote by $t^{lmn}_{\alpha,\beta}$ the hopping amplitudes along the connecting direction l$\mathbf{x}$ + m$\mathbf{y}$ + n$\mathbf{z}$  between the orbitals $\alpha$ and $\beta$.
We indicate also the $\sigma$ and $\pi$ chemical bonds. We used the DFT values as reported in Ref. \onlinecite{Islam19} for all the parameters except than $V^{110}_{\sigma cc}$. We only varied $V^{110}_{\sigma cc}$, we change this value depending on whether we want to reproduce the experimental band gap \cite{Dornhaus83} or the Chern number.
We used the same procedure for PbSe, PbTe and SnTe. 
The decision to modify $V^{110}_{\sigma cc}$ is due to the fact that we cut all the hoppings beyond the second nearest neighbours, the larger parameters we cut concern the cation-cation hybridization, so in this way we balanced the contribution due to the cations and the anions.

In the next three subsections, we report three tight-binding minimal models and the associated Chern number map for the different cases. In the last subsection, we add the Weyl phase to obtain the full topological phase diagram.

\subsection{Tight-binding model to reproduce the experimental band gap of PbSe and PbTe and the theoretical band gap of SnTe}

\begin{table} [!]
\begin{center}
\begin{tabular}{ |c||c||c|c|c|  }
 \hline
& & SnTe & PbTe  & PbSe \\
 \hline
\multirow{2}{4em}{On-site} & $\epsilon_{c}$  &  1094.6     & 1855.4  &  2250.4  \\
& $\epsilon_{a}$  & -1205.4  & -646.0    & -1206.1   \\
\hline
 \multirow{2}{4em}{SOC} & $\lambda_{c}$ & 334.5 & 1068.2  & 1132.5  \\
&  $\lambda_{a}$    & 524.4 & 553.8 &  250.3  \\
 \hline
 \multirow{2}{4em}{cation-anion} &   $V^{100}_{\sigma}$   & 1916.1 & 1807.4 & 1788.9\\
 & $V^{100}_{\pi}$  & -434.2   & -402.6 & -353.6\\
  \hline
 \multirow{2}{4em}{cation-cation} & $V^{110}_{\sigma cc}$  &     142.9 & 193.3   & 465.21 \\
 & $V^{110}_{\pi cc}$  &     -25.3 & -23.8   & 21.1 \\
 \hline
 \multirow{2}{4em}{anion-anion} 
& $V^{110}_{\sigma aa}$  &     -200.8 & -90.9   & -108.2 \\
 & $V^{110}_{\pi aa}$  & 248.6   &     131.5 & 204.0    \\
 \hline
\end{tabular}
\end{center}
\caption{Values of the electronic parameters of our tight-binding model for SnTe, PbTe and PbSe. The unit is meV. This hamiltonian reproduces the experimental band gap at the L point and at T=0 K for PbTe \cite{Dornhaus83} and PbSe \cite{Dornhaus83} and the theoretical band gap for SnTe \cite{hsieh2012topological}. }
 \label{tabhoppings1}
\end{table} 

\begin{table} [!]
\begin{center}
\begin{tabular}{ |c||c||c|c|c|  }
 \hline
& & SnTe & PbTe  & PbSe \\
 \hline
\multirow{2}{4em}{On-site} & $\epsilon_{c}$  &  1094.6     & 1855.4  &  2250.4  \\
& $\epsilon_{a}$  & -1205.4  & -646.0    & -1206.1   \\
\hline
 \multirow{2}{4em}{SOC} & $\lambda_{c}$ & 334.5 & 1068.2  & 1132.5  \\
&  $\lambda_{a}$    & 524.4 & 553.8 &  250.3  \\
 \hline
 \multirow{2}{4em}{cation-anion} &   $V^{100}_{\sigma}$   & 1916.1 & 1807.4 & 1788.9\\
 & $V^{100}_{\pi}$  & -434.2   & -402.6 & -353.6\\
  \hline
 \multirow{2}{4em}{cation-cation} & $V^{110}_{\sigma cc}$  &     150.0 & 202.9   & 488.5 \\
 & $V^{110}_{\pi cc}$  &     -25.3 & -23.8   & 21.1 \\
 \hline
 \multirow{2}{4em}{anion-anion} 
& $V^{110}_{\sigma aa}$  &     -200.8 & -90.9   & -108.2 \\
 & $V^{110}_{\pi aa}$  & 248.6   &     131.5 & 204.0    \\
 \hline
\end{tabular}
\end{center}
\caption{Same as in Tab. \ref{tabhoppings2} but modifying $V^{110}_{\sigma cc}$ in  $V^{110}_{\sigma cc}$ (1 + $\alpha$), with $\alpha$=0.05.  The unit is meV. This hamiltonian reproduces reasonably well the experimental topology as shown in  Fig. \ref{Chern2}}
 \label{tabhoppings2}
\end{table} 

\begin{table} [!]
\begin{center}
\begin{tabular}{ |c||c||c|c|c|  }
 \hline
& & SnTe & PbTe  & PbSe \\
 \hline
\multirow{2}{4em}{On-site} & $\epsilon_{c}$  &  1094.6     & 1855.4  &  2250.4  \\
& $\epsilon_{a}$  & -1205.4  & -646.0    & -1206.1   \\
\hline
 \multirow{2}{4em}{SOC} & $\lambda_{c}$ & 497.3 & 1068.2  & 1132.5  \\
&  $\lambda_{a}$    & 581.8 & 553.8 &  250.3  \\
 \hline
 \multirow{2}{4em}{cation-anion} &   $V^{100}_{\sigma}$   & 1916.1 & 1807.4 & 1788.9\\
 & $V^{100}_{\pi}$  & -434.2   & -402.6 & -353.6\\
  \hline
 \multirow{2}{4em}{cation-cation} & $V^{110}_{\sigma cc}$  &     213.4 & 193.3   &  465.21 \\
 & $V^{110}_{\pi cc}$  &     -25.3 & -23.8   & 21.1 \\
 \hline
 \multirow{2}{4em}{anion-anion} 
& $V^{110}_{\sigma aa}$  &     -200.8 & -90.9   & -108.2 \\
 & $V^{110}_{\pi aa}$  & 248.6   &     131.5 & 204.0    \\
 \hline
\end{tabular}
\end{center}
\caption{Values of the electronic parameters of our tight-binding model for SnTe, PbTe and PbSe. The unit is meV. With this hamiltonian, we reproduce the experimental band gap at the L point and at T=0 K for PbTe \cite{Dornhaus83} and PbSe \cite{Dornhaus83}, and we obtain a good approximation of the experimental band gap for SnTe \cite{Dornhaus83}.}
 \label{tabhoppings3}
\end{table} 

\begin{figure}[th]
	\centering
	\includegraphics[scale=0.22]{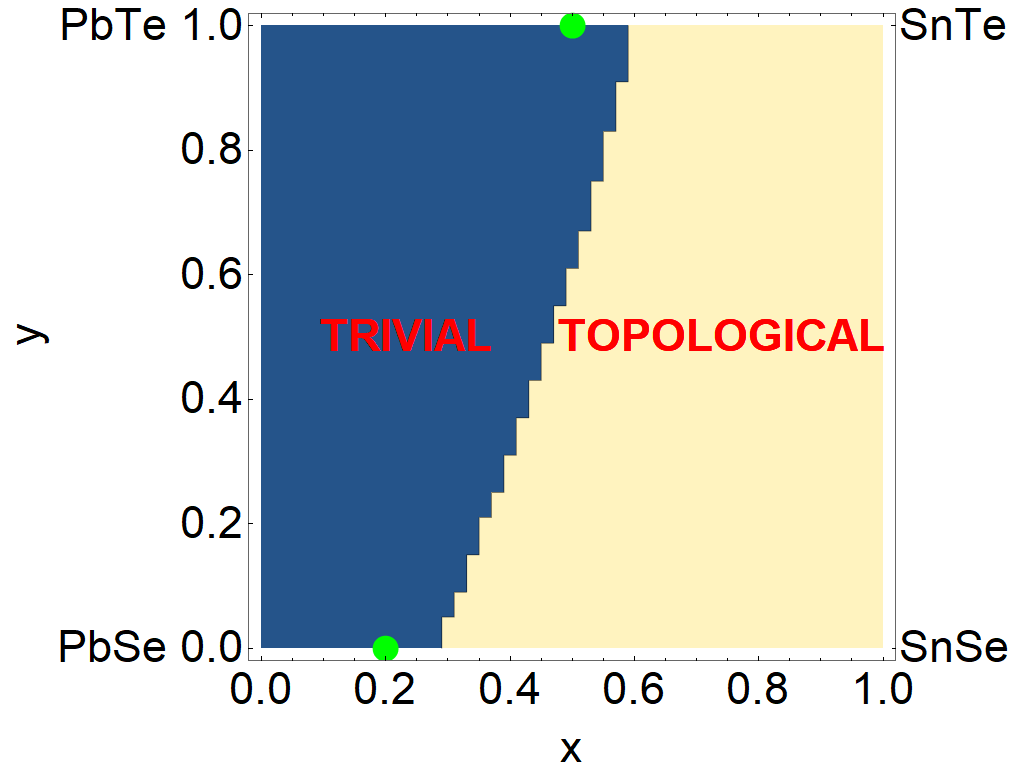}
	\caption{Mirror Chern Number for the quaternary compound Pb$_{1-x}$Sn$_x$Se$_{1-y}$Te$_y$ as a function of the concentrations x and y obtained with the hamiltonian reported in Tab. \ref{tabhoppings1}. With this hamiltonian, we reproduce the experimental band gap at the L point and at T=0 K for PbTe \cite{Dornhaus83} and PbSe \cite{Dornhaus83} and the theoretical band gap for SnTe \cite{hsieh2012topological}. The green circles are the experimental transition points \cite{Dornhaus83}. The blue region represents the trivial phase, while the yellow region represents the topological phase.	
	\label{Chern1}}
	\end{figure}

\begin{figure}[th]
	\centering
	\includegraphics[scale=0.22]{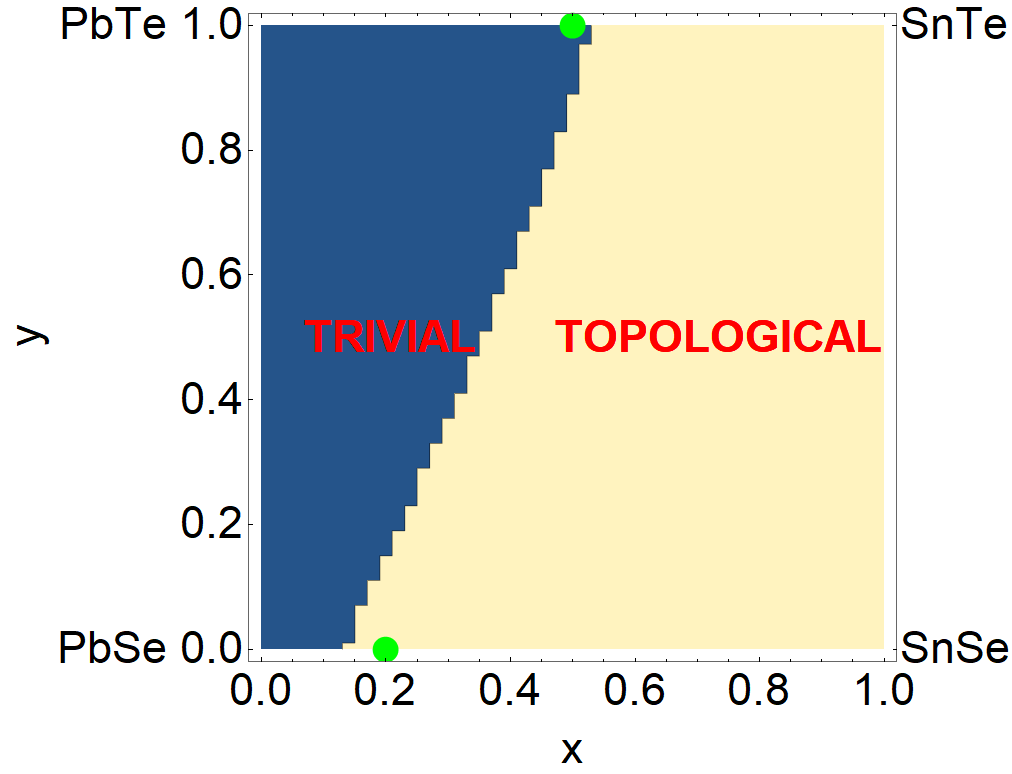}
	\caption{Same as in Fig. \ref{Chern1}, but obtained with the hamiltonian reported in Tab. \ref{tabhoppings2} and that reproduces the closure of the trivial gap. 
	}
	\label{Chern2}
\end{figure}

\begin{figure}[th]
	\centering
	\includegraphics[scale=0.22]{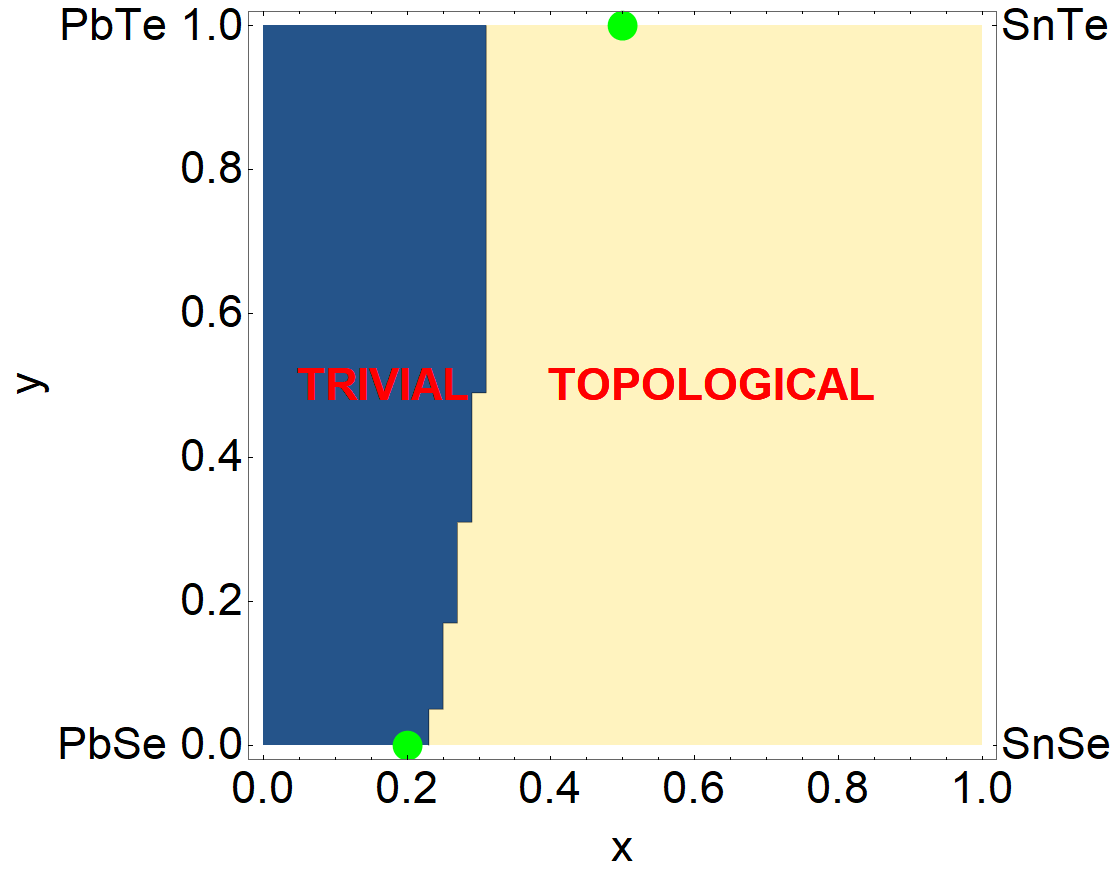}
	\caption{Same as in Fig. \ref{Chern1}, but obtained with the hamiltonian reported in  Tab. \ref{tabhoppings3} and that reproduces the experimental band gap at the L point and at T=0 K for PbSe, PbTe and SnTe \cite{Dornhaus83}.
	} 	
	\label{Chern3}
\end{figure}

The map obtained with the hoppings reported in Tab. \ref{tabhoppings1} is shown in Fig. \ref{Chern1}, where the  experimental transition points are also indicated.
The border that separates the trivial and the topological regions is not linear, differently from what one could expect from the VCA approximation that reproduces a linear combination of the Hamiltonian.
We obtain a curve quite close to the experimental points with these values of the hopping parameters.
With the hamiltonian of Table \ref{tabhoppings1}, we reproduce the experimental band gap at the L point and at T=0 K for PbTe \cite{Dornhaus83} and PbSe \cite{Dornhaus83} and the theoretical band gap for SnTe \cite{hsieh2012topological}, which is -0.185 eV.
\\

\subsection{Tight-binding model to reproduce the topology}

 To improve the agreement with the experimental points in Fig. \ref{Chern1}, we act on $V^{110}_{\sigma cc}$ introducing the parameter $\alpha$, such as the new hybridization will be $V^{110}_{\sigma cc}$ (1 + $\alpha$).
In Tab. \ref{tabhoppings2} the case with $\alpha$=0.05 is reported, and in Fig. \ref{Chern2} the map obtained with this hamiltonian is shown. 
In this case, we do not reproduce the band gap but we can see from Fig. \ref{Chern2} that the border between the trivial and the topological phase is closer to the experimental points. With this choice of the parameter $\alpha$ we increased the width of the topological region; we verified that if we choose a negative value of the $\alpha$ parameter we go in the other direction increasing the trivial region of the map.
This model reproduces well the closure of the trivial band gap.

\subsection{Tight-binding model to reproduce the experimental band gaps}

Furthermore, in Tab. \ref{tabhoppings3}  we report the hamiltonian which reproduces the experimental band gap of PbSe\cite{Dornhaus83} and PbTe \cite{Dornhaus83} at the L point and at T=0 K and we obtain a good approximation of the band gap of SnTe \cite{Dornhaus83} at T=0 K. With this hamiltonian the band gap obtained is 0.145 eV for PbSe, 0.190 eV for PbTe and $\sim$ $-$0.321 eV for SnTe. In this case, we varied not only $V^{110}_{\sigma cc}$ but also the SOC coupling constants respect to the other Tables.
The Chern map obtained with this hamiltonian is reported in Fig. \ref{Chern3}.

We have tuned the spin-orbit to reproduce the band gap of the pure phases, however, we lose accuracy on the topological transition.
Therefore, within this simplified model, we were not able to accurately reproduce 
both topological transition and band gap of the pure phases.

\subsection{Weyl phase}

\begin{figure}[th]
	\centering
	\includegraphics[scale=0.22]{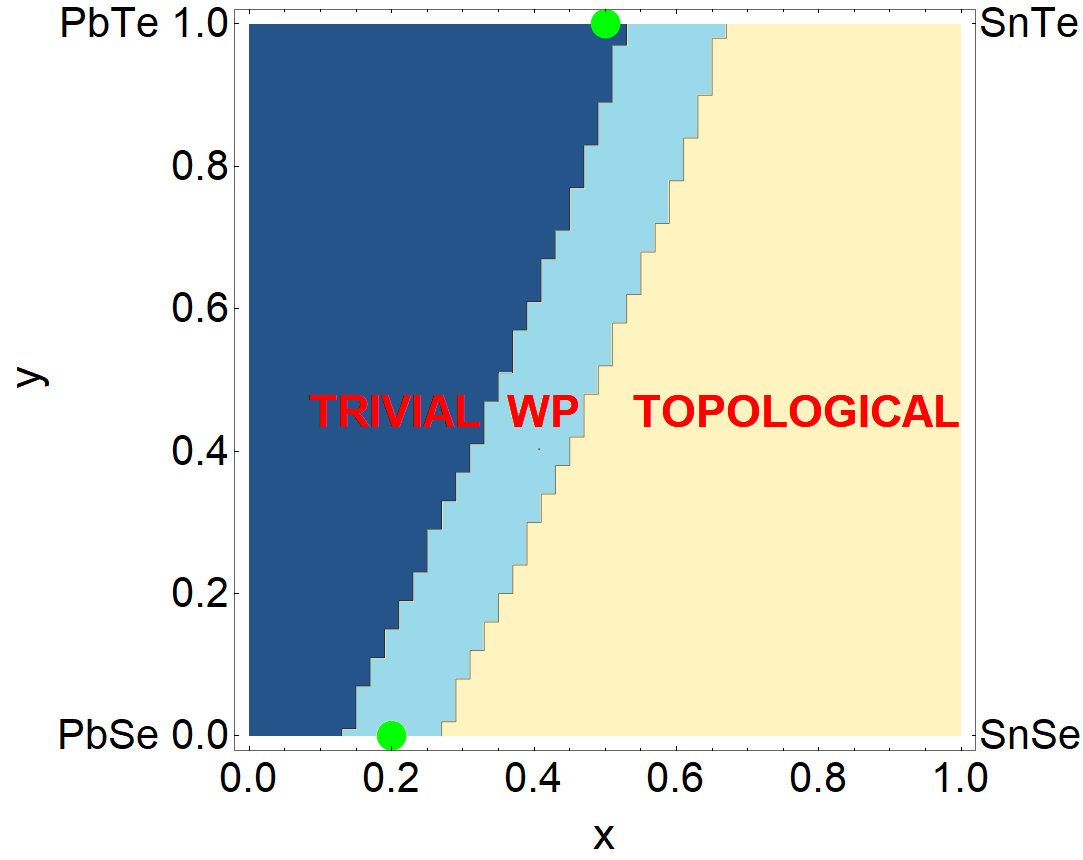}
	\caption{Complete topological phase diagram using the hamiltonian of Table \ref{tabhoppings2} with the topological phase divided in Weyl semimetallic phase in light blue and topological insulating phase in yellow according to the results reported by {\L}usakowski et al\cite{Lusakowski18}.
	\label{Weyl}}
\end{figure}

Recently, it has been shown that the transition between the topological and the trivial phase is broadened \cite{Lusakowski18,Wang19b}. The region where the gap is zero is not a single point but it is wider, the width is of the 15 \% of the concentration starting from the point where the gap gets closed.
A WP has been found in the zero gap region, where Wang et al.\cite{Wang19b} observe four Weyl points forming
two Weyl pairs in the alloy (PbSe)$_{1-x}$(SnSe)$_x$ by them investigated. They show that the sequential band inversion regime always has non-zero separation between the Weyl pairs. Therefore, the WPs exist not accidentally but in a wide range within the zero band gap range. This WP is present both if we investigate the alloy under pressure or as a function of concentration.  
We also know that the VCA is valid close to x=0 and x=1, and which it is less reliable in the intermediate region. The region where the VCA does not work well, namely the intermediate region between 0 and 1, coincides with the region of the closure of the band gap, where the Weyl phase is present. Because the width of this phase is of the 15$\%$ of the concentration starting from the region where the gap gets closed, we can place this phase in the map based on experimental data. In the other regions, namely close to 0 and 1, the VCA works well therefore our Chern map is reliable.
In Fig. \ref{Weyl} we report the mirror Chern number map obtained with the hamiltonian of Table \ref{tabhoppings2}, by showing not only the transition line between the topological insulating phase and TCI phase, but also the WP region, which is located in an area of 15\% of width starting from the point of the closing of the trivial gap \cite{Lusakowski18,Wang19b}.

\section{Conclusions}
We have built different minimal tight-binding models for the three compounds PbSe, PbTe and SnTe in order to investigate the topological phase diagram of the quaternary compound Pb$_{1-x}$Sn$_x$Se$_{1-y}$Te$_y$.
We analyzed the properties of the quaternary compound Pb$_{1-x}$Sn$_x$Se$_{1-y}$Te$_y$ for the first time, since previously only the ternary compounds Pb$_{1-x}$Sn$_x$Te and Pb$_{1-x}$Sn$_x$Se have been studied. The transition between the trivial and the topological phases driven by the x content of Sn was investigated in previous works, but an entire phase map for these compounds had not yet been provided. We built up a model that can reproduce the whole Chern map for the quaternary compound and we give indications to the experimentalists about the border between the trivial and the topological phases. An experimental transition line between trivial and topological phases is not available because only the ternary compounds were studied, but with one of our sets of hopping parameters we obtain a good agreement with the experimental points related to the ternary compounds Pb$_{1-x}$Sn$_x$Te and Pb$_{1-x}$Sn$_x$Se. Another important information that we get is that the border between the trivial and the topological regions is not linear respect to what one can expect if virtual crystalline approximation is used. 
Furthermore, since we know from previous studies that the region where the gap is closed is not a single point but wider and the width is of the 15$\%$ of the concentration starting from the point where the gap gets closed, we give an indication about the position of this region in the Chern map.
Once obtained the tight-binding parameters extracted from {\it ab-initio}, we tuned the parameter $V^{110}_{\sigma cc}$ to improve the agreement with the transition of the Mirror Chern number at y=0 and y=1 observed in the experimental results. Once, we satisfactory reproduce the topological transition line close to the experimental transition points at y=0 and y=1, and we also add the location of the Weyl phase predicted in the literature\cite{Wang19b} producing the full topological phase diagram for Pb$_{1-x}$Sn$_x$Se$_{1-y}$Te$_y$ quaternary compound. \\

\section{Aknowledgments}
We thank V. Volobuiev, R. Buczko and T. Hyart for useful discussions.
The work is supported by the Foundation for Polish Science through the International Research Agendas program co-financed by the
European Union within the Smart Growth Operational Programme.
We acknowledge the access to the computing facilities of the Interdisciplinary
Center of Modeling at the University of Warsaw, Grant G84-0, GB84-1 and GB84-7.
We acknowledge the CINECA award under the ISCRA initiative  IsC85 "TOPMOST" and IsC93 "RATIO" grant, for the availability of high-performance computing
resources and support.

\newpage
\medskip
\bibliography{SnTe_Carmine}

\end{document}